\documentclass[structabstract]{aa}  
\usepackage{multirow}
\usepackage{graphicx}
\usepackage{txfonts}
\usepackage{natbib}
\def\lesssim{\mathrel{\hbox{\rlap{\hbox{\lower4pt\hbox{$\sim$}}}\hbox{$<$}}}}
\def\gtrsim{\mathrel{\hbox{\rlap{\hbox{\lower4pt\hbox{$\sim$}}}\hbox{$>$}}}}

\begin{document}

\title{Comparing the asteroseismic properties of pulsating pre-extremely low mass white dwarf and $\delta$ Scuti stars} 
 
\author{Julieta P. S\'anchez Arias\inst{1,2}, Alejandra D. Romero\inst{3}, Alejandro H. C\'orsico\inst{1,2},  Ingrid Pelisoli\inst{3}, Victoria Antoci\inst{4}, S. O. Kepler\inst{3}, Leandro G. Althaus\inst{1,2} \and Mariela A. Corti\inst{1,5}}
\institute{$^{1}$ Facultad de Ciencias Astron\'omicas y Geof\'{\i}sicas, 
          Universidad Nacional de La Plata, Paseo del Bosque s/n, 1900 
          La Plata, Argentina\\
          $^{2}$ Instituto de Astrof\'isica La Plata, CONICET-UNLP, Argentina\\
          $^{3}$ Instituto de F\'sica, Universidade Federal do Rio Grande do Sul, 91501-900 Porto-Alegre, RS, Brazil.\\
         $^{4}$ Stellar Astrophysics Centre, Aarhus University, Ny Munkegade 120, DK-8000 Aarhus C, Denmark \\ 
       $^{5}$ Instituto Argentino de Radioastronomía, CCT-La Plata, CONICET, C.C. Nro. 5, 1984 Villa Elisa, Argentina\\        \email{jsanchez@fcaglp.unlp.edu.ar}     
           }
\date{Received ; accepted }

\abstract {Pulsating extremely low-mass pre-white dwarf stars (pre-ELMV), with masses between $\sim 0.15 M_{\sun}$ and $\sim 0.30 M_{\sun}$, constitute a new class of variable stars showing $g$- and possibly $p$-mode 
pulsations with periods between $320$ and $6000$ s (frequencies between 14.4 and 270 c/d), driven by the $\kappa$ mechanism operating in the second He ionization zone. On the other hand, main sequence $\delta$ Scuti 
stars, with masses between $1.2-2.5 M_{\sun}$, pulsate in low-order $g$ and $p$ modes with periods in the range $[700-28\,800]$ s (frequencies in the range [3-123] c/d), driven by the $\kappa$ mechanism operating in 
the He II ionization zone and the turbulent pressure acting in the HI ionization layer. Interestingly enough, the instability strips of pre-ELM white dwarf and $\delta$ Scuti stars nearly overlap in the 
$T_{\rm eff}$ vs. $\log g$ diagram, leading to a degeneracy when spectroscopy is the only tool to classify the stars and pulsation periods only are considered.}
{Pre-ELM white dwarf and $\delta$ Scuti stars are in very different stages of evolution and therefore their internal structure is very distinct. This is mirrored in their pulsational behavior, thus employing asteroseismology 
should allow us to distinguish between these groups of stars despite their similar atmospheric parameters.}
{We have employed adiabatic and non-adiabatic pulsation spectra for models of pre-ELM white dwarfs and $\delta$ Scuti stars, and compare their pulsation periods, period spacings, and rates of period change. }
{Unsurprisingly, we found substantial differences in the period spacing of $\delta$ Scuti and pre-ELM white dwarf models. Even when the same period range is observed in both classes of pulsating stars, the modes 
have distinctive signature in the period spacing and period difference values. For instance, the mean period difference of $p-$ modes of consecutive radial orders for $\delta$ Scuti model are at least four times 
longer than the mean period spacing for the pre-ELM white dwarf model in the period range $[2\,000 - 4\, 600]$ s (frequency range $[18.78-43.6]$ c/d). In addition, the rate of period change is two orders of magnitudes 
larger for the pre-ELM white dwarfs compared to $\delta$ Scuti stars. In addition, we also report the discovery of a new variable star, SDSSJ075738.94$+$144827.50, located in the region of the $T_{\rm eff}$ vs. $\log g$
diagram where these two kind of stars coexist.}
{The characteristic spacing between modes of consecutive radial orders ($p$ as well as $g$ modes) and the large differences found in the rates of period change for $\delta$ Scuti and pre-ELM white dwarf stars suggest 
that asteroseismology can be employed to discriminate between these two groups of variable stars. Furthermore, we found that SDSSJ075738.94$+$144827.50 exhibits a period difference between $p$ modes characteristic of a 
$\delta$ Sct star, assuming consecutive radial order for the observed periods.}
\authorrunning{S\'anchez Arias et al.}
\titlerunning{$\delta$ Scuti and pre-ELMV WD stars}
\maketitle

\section{Introduction}
\label{introduction}

A significant number of pulsating stars were discovered by the COnvection ROtation and planetary Transits (COROT) satellite \citep{2009A&A...506..411A}, the NASA's Kepler space telescope \citep{2016RPPh...79c6901B}, and by the Kepler's second mission: K2 \citep{2016PASP..128g5002V}. Moreover, the upcoming NASA Transiting Exoplanet Survey Satellite (TESS) will provide excellent photometric precision and long intervals of uninterrupted observations of of almost the entire sky \citep{2015JATIS...1a4003R}. These observations will rapidly increase the population of the Hertzsprung--Russell (HR) diagram with new families of variable stars and therefore a proper classification of the observed objects is essential. In particular, there is a region in the $T_{\rm eff}-\log g$ diagram which is being particularly rapidly populated and in which two very different families of variable stars coexists: main sequence (MS) $\delta$ Scuti ($\delta$ Sct) stars and the pulsating extremely low-mass pre-white dwarf stars (pre-ELMV). These two families of variable stars share very similar atmospheric characteristics, which makes it difficult to distinguish them from spectroscopy alone. The main goal of this work is to provide asteroseismic tools to enable their correct classification and to increase the number of known pre-ELMVs in order to study their very interesting evolution and stellar structure.

Pulsating stars in the lower part of the classical instability strip are associated with early stages of evolution, the pre-MS, MS, and the immediate post-MS. The most numerous pulsator class among them are known as $\delta$ Sct stars showing low to intermediate radial order pressure, mixed and gravity modes. 

In smaller numbers, we also have other types of pulsators such as the $\gamma$ Doradus ($\gamma$ Dor) stars, that pulsate in high radial order gravity modes. Because the instability strip of $\gamma$ Dor and $\delta$ Sct stars overlap, we can also find stars showing both $\delta$ Sct and $\gamma$ Dor pulsations, also known as hybrid pulsators. A third group of pulsators in this region are the rapid oscillating (roAp) stars, which show periods on the order of ten minutes and display strong magnetic fields.

Recently, two new classes of pulsating white dwarfs (WD) stars were found in a similar region of the $T_{\rm eff}-\log g$ diagram, known as extremely low mass variable (ELMV) stars and their evolutionary precursors, the pre-ELMV stars. The ELM WD stars are characterized by stellar masses $\lesssim 0.3 M_{\sun}$ \citep{2010ApJ...723.1072B}, and are thought to be the result of strong mass--transfer events at the red giant stage of low-mass stars evolution in close binary systems \citep[see e.g.,][]{2014A&A...571A..45I,2016A&A...595A..35I,2013A&A...557A..19A}. A single star of such a low mass would take more than the age of the Universe to evolve into the WD stage clearly implying binary evolution. In 2012, the first ELMV WD star was discovered \citep{2012ApJ...750L..28H}, opening the prospect for probing the internal structure of this kind of stars. 

Pre-ELMV stars are thought to be the progenitors of the ELMV stars. The pre-ELMVs show pulsations with periods compatible with low order radial and non-radial $p$ and $g$ modes. The excitation mechanism is the $\kappa-\gamma$ mechanism acting mainly in the zone of the second He ionization (HeII layer) \citep{2016A&A...588A..74C,2016ApJ...822L..27G,2016A&A...595A..35I}. Their effective temperatures are similar to those of ELMVs, but they are less compact, showing lower surface gravities. The location of the pre-ELMV WDs in the $T_{\rm eff}$--$\log g$ diagram is one of our main interests in this paper, since it partially overlaps with the classical instability strip.

Observations show that not only the instability strip for both $\delta$ Sct stars and pulsating pre-ELM WDs overlap in the $T_{\rm eff}-\log g$ plane, but they also share a common range of excited periods, thus making the correct identification of each class more difficult.  One way to distinguish the $\delta$ Sct from pre-ELM WD stars is by knowing their mass, of course. As mentioned, pre-ELM WD stars have stellar masses $\lesssim 0.3 M_{\sun}$ while $\delta$ Sct stars have masses between $1.5-2.5 M_{\sun}$. Also, the stellar radius characteristic of a MS star is two to three times the radius of a pre-ELM WD star. The surface gravity depends on both the stellar mass and radius, there is a degeneracy in the surface gravity, which means we cannot determine both parameters independently using only spectroscopy. However, we can estimate the radius of the star if we know its distance. The Gaia satellite will measure parallaxes, and thus distances, for most of the stars in the Milky Way, giving rise to a very detail map of our Galaxy \citep{2016A&A...595A...1G,2016A&A...595A...2G}. The first data release was published by \citet{2016A&A...595A...2G}, consisting of astrometry and photometry for over one billion sources brighter than magnitude 20.7, and DR2 is scheduled for April 2018. However, most of the ELM and pre-ELM WD stars are in binary systems and are usually found by looking for high proper motion objects, since they are intrinsically faint and must be close by to be observed \citep{2017ASPC..509..447P}. In addition, approximately $50 \%$ of  $\delta$ Sct stars, can be found in binary systems as well. A more complex data reduction is necessary for some of this objects with high proper motions, in binary systems and showing photometric variabilities and therefore most of the parallax and distances measures will only be available after Gaia finishes with the observations. This shows the need to find other tools to correctly separate MS $\delta$ Sct from pre-ELM variable stars that do not rely on spectroscopy or parallax measures.

In this work we present such tools, by taking advantage of the pulsating nature of MS $\delta$ Sct and pre-ELM variable stars. We address a comparison of the pulsational properties of both MS $\delta$ Sct stars and pre-ELM WD stars, aiming at providing asteroseismic tools for their correct classification. Specifically, we present the period and frequency spacings ($\Delta \Pi= \Pi_{k+1}-\Pi_{k}$ and $\Delta \nu=\nu_{k+1}-\nu_{k}$, respectively, $k$ being the radial order) and the rate of the period changes ($d\Pi/dt$) as tools to distinguish MS $\delta$ Sct stars from pre-ELMV WDs. The differences expected in these two quantities allow us to shed light on the nature of both classes of stars and to classify them without the knowledge of their masses and radii. To this end, we have considered two different regions in the $T_{\rm eff}-\log g$ plane where pulsating objects are currently observed and can be confused with $\delta$ Sct or pre-ELMV stars. A cool region with a characteristic effective temperature of $\delta$ Sct stars ($T_{\rm eff} \sim 7500$~K) where pre-ELMV are predicted to lie and a hot region at $T_{\rm eff} \sim 9600$ K where $\delta$ Sct and pre-ELMV stars can be confused if we consider the external uncertainties in $\log g$ ($~ 0.25$ dex) and $T_{ \rm eff}$ (up to 5$\%$) of the pulsating objects found in this region \citep{2016MNRAS.455.3413K,2018MNRAS.475.2480P}. For each region, we have analyzed the pulsational properties of stellar models for MS $\delta$ Sct stars computed by  \citet{2017A&A...597A..29S} and pre-ELM WD models computed by \citet{2016A&A...588A..74C}. All stellar models are computed using the stellar evolutionary code LPCODE \citep{Althausetal2005,Althausetal2009,Althausetal2013}, that calculates the full evolution of low- and intermediate-mass stars from the zero age main sequence (ZAMS). For each stellar model, we analyzed the period range of adiabatic pulsation modes, the period and frequency spacings, and the evolutionary rates of period change for $p$ and $g$ modes, and also study the pulsational instability through non-adiabatic pulsation computations.  

The paper is organized as follows. In Section \ref{observed} we describe in detail the observational properties of the variables in lower part of the classical instability strip and pre-ELMV pulsators. In Section \ref{numerical-tools} we describe the numerical tools employed in building the  models considered in our analysis. Section \ref{asteroseismic-comparison} is devoted to the analysis and comparison of the pulsational properties of MS $\delta$ Sct (hereinafter $\delta$ Sct) and pre-ELMV stars, highlighting their similarities and differences. The non-adiabatic pulsational properties, in particular the instability of the pulsation modes, is presented in Section \ref{no-adiabatico}. In Section \ref{testing} we apply the analysis described in Section \ref{asteroseismic-comparison} to a new pulsating star, J075739.94$+$144827.5 and the stars J173001.94+070600.25 and J145847.02+070754.46 reported in \citet{2016A&A...587L...5C}. Finally, we summarize our results in Section \ref{conclusions}.

\section{Observational properties}
\label{observed}

The $\delta$ Sct stars are variables of spectral type between A3 and F5, lying on the extension of the Cepheid instability strip toward effective temperatures between $6\, 700$ K and $8\, 500$ K and with stellar masses in the interval $1.2-2.5 M_{\sun}$. Pulsation periods can be between $\sim 700\ {\rm s}$ to $\sim 28\,800$ s (frequencies between 3 c/d - 123 c/d) (Table \ref{resumenperiod}) \citep[see, e.g.,][]{2012ApJ...759...62B,2011A&A...534A.125U,2010AN....331..989G}, corresponding to radial and non-radial low and intermediate order $p$ and $g$ modes, mainly driven by the $\kappa$ mechanism operating in the He II partial ionization zone \citep[see, e.g.,][]{2005A&A...435..927D} and the turbulent pressure acting in the hydrogen ionization zone \citep{2014ApJ...796..118A}. $\delta$ Sct stars are subdivided into many categories \citep[e.g.,][]{2000BaltA...9..149B}, including the sub-classes of Population I and II stars. The latter are also known as SX Phe stars, they are usually blue straggler stars presumably formed by the merger of two MS stars \citep{2010aste.book.....A} and their belonging to the MS is today an open question. SX Phe stars are usually connected with low metallicities. However, recently \citet{2016arXiv161110332N} have discovered SX Phe stars with near solar metallicity.

The $\gamma$ Dor stars are variable stars with spectral types of A7-F5 located near the cool border of the classical instability strip, encompassing regions in the HR diagram with effective temperatures between 6900 K and 7700 K and stellar masses in the interval $1.3-1.8 M_{\sun}$. They pulsate in non-radial high order $g$ modes with periods between $26\,000$ s and $260\,000$ s (frequencies between 0.3 c/d - 3.3 c/d), reaching up to $12\,342$ s (7 c/d) when rapid rotation is involved \citep[see, e.g.,][]{2013MNRAS.429.2500B,2017MNRAS.465.2294O}. These modes are driven by a convective flux blocking mechanism operating in the base of their convective envelope \citep{2000ApJ...542L..57G}. 

The instability strips of $\delta$ Sct and $\gamma$ Dor stars partially overlap in a region in which we find the so-called hybrid $\delta$ Sct-$\gamma$ Dor stars. These stars show pulsations typical for both $\delta$ Sct stars and $g$-mode variables \citep[see, e.g.,][]{2017A&A...597A..29S}.

The roAp stars are also located close to the MS where $\delta$ Sct lies. They are strongly magnetic Population I stars with a peculiar chemical surface composition caused by atomic diffusion. They exhibit high-order low-degree $p$ modes pulsations, with periods in the range 339-$1\,272$ s (frequencies in the range 67-254 c/d) \citep{2010aste.book.....A}.

\begin{figure*} 
\begin{center}
\includegraphics[clip,width=18 cm]{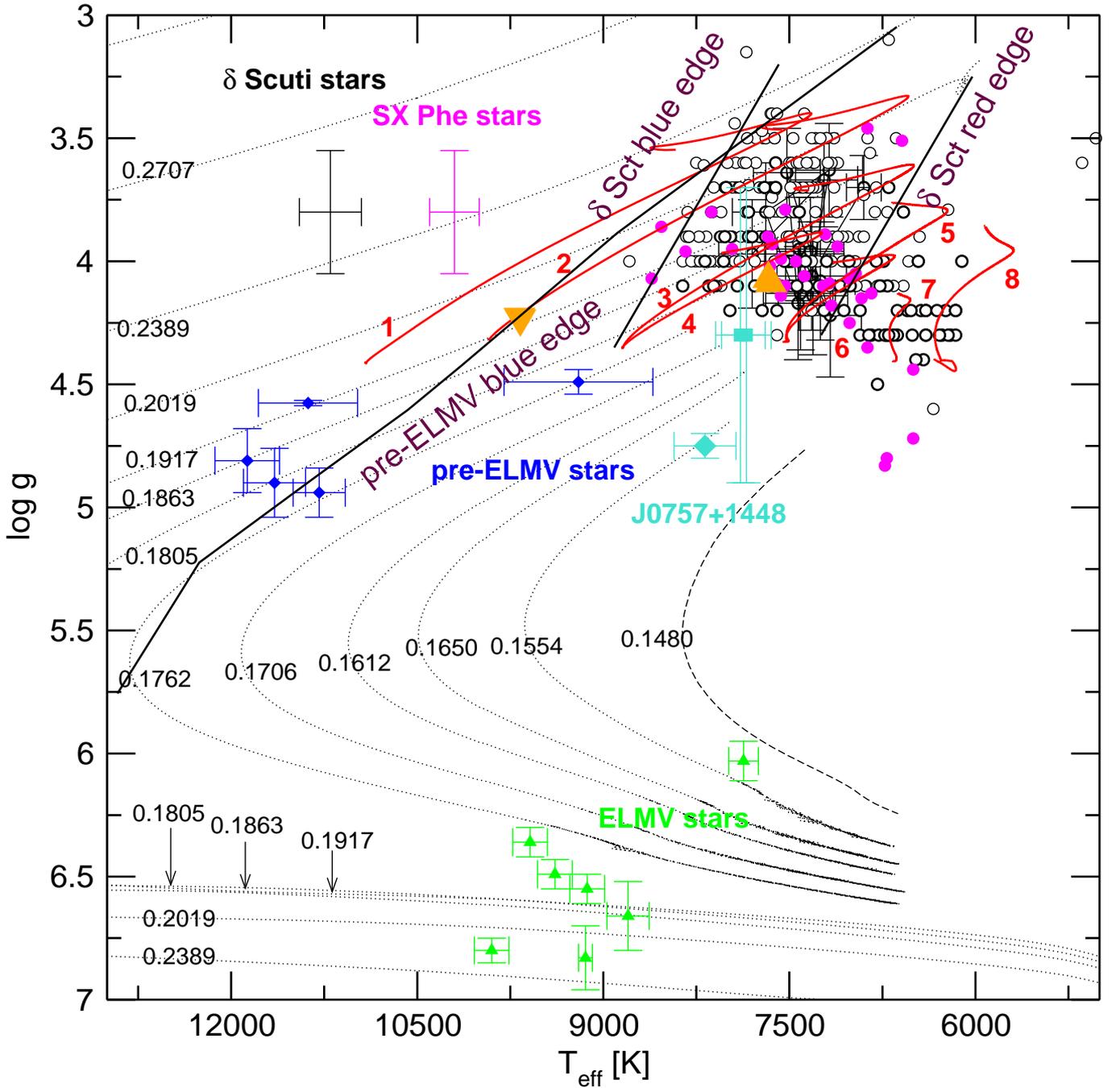} 
\caption{$T_{\rm eff}-\log g$ diagram showing the location of ELMV stars (light green triangles), pre-ELMV stars (blue diamonds), $\delta$ Sct stars (white circles) including SX Phe stars (magenta circles). The atmospheric parameters are extracted from different works detailed in the main text. Also, we included the theoretical evolutionary tracks of low mass He--core WDs (black dotted and dashed lines) from \citet{2013A&A...557A..19A} and \citet{2001MNRAS.325..607S}, and of MS evolutionary tracks (red lines) from \citet{2017A&A...597A..29S}. Black numbers correspond to different values of the stellar mass of the low mass He--core WDs, whereas red numbers are associated to the values of the mass, the metallicity and the overshooting parameter (see Table \ref{msmodels}). Orange triangles show the position of the template models (see Section \ref{asteroseismic-comparison}). The cyan squares indicate the position of the two stars reported by \citet{2016A&A...587L...5C}, J1730+0706 and J1458+0707, and the cyan diamond represents the position of a new pulsating star, J0757+1448. These three objects will be analyzed in Section \ref{testing}.}
\label{HRcompleto} 
\end{center}
\end{figure*}

The number of ELM WDs has increased in recent years thanks to large scale surveys, like the Sloan Digital Sky Survey \citep{2010ApJ...723.1072B, 2016arXiv161206390B}. Pulsations in some ELM WDs have been reported from $2012$ onward \citep{2012ApJ...750L..28H,2013ApJ...765..102H, 2013MNRAS.436.3573H,  2015MNRAS.446L..26K, 2016arXiv161206390B}. 
They oscillate in non-radial $g$ modes, with periods typically in the range $[1000-6\,300]$ s \citep{2017ASPC..509..289C} and some of them may also exhibit $p$ modes, which are not common to observe in a WD star \citep{2013ApJ...765..102H}. Their effective temperatures and surface gravities are in the ranges $7\,000 \lesssim T_{\rm eff} \lesssim 10\,000$ K and $6 \lesssim \log g \lesssim 7$ respectively.

To date, five objects identified as variable pre--ELM WD stars are known \citep{2016ApJ...822L..27G, 2013Natur.498..463M, 2014MNRAS.444..208M}. They are characterized by surface gravities and effective temperatures in the theoretical ranges $3 \lesssim 
\log g \lesssim 5.5$ and $6\,000 \lesssim T_{\rm eff} \lesssim 12\,000$ K respectively, a region  of the $\log g-T_{\rm eff}$ plane that partially overlaps with the region where some variable stars in the classical instability strip are usually found. The excitation mechanism is the $\kappa-\gamma$ mechanism acting in the HII ionization zone \citep{2016A&A...588A..74C,2016ApJ...822L..27G,2016A&A...595L..12I}. According to \cite{2016ApJ...822L..27G}, the pre-ELMV objects J0756$+$6704, J1141$+$3850 and J1157$+$0546 are classified as $p$-mode pulsators with periods between 320 and 590 s. On the other hand, the other two pre-ELMVs known, WASP J0247$-$25B and WASP J1628$+$10B reported in \citet{2013Natur.498..463M, 2014MNRAS.444..208M} seem to have a non-radial and radial mixed modes. Besides these well-known five pre-ELMVs, there are three more objects reported in \citet{2016A&A...587L...5C} and \citet{2016ApJ...821L..32Z} as possible pre-ELMV with periods that reaches $6\,000$ s (frequencies above 14.4 c/d).

In Tables \ref{resumenspectro} and \ref{resumenperiod} we summarize the atmosphere parameters and observed period and frequency range of the different classes of variable stars described in this section.

\begin{table}[h]
  \centering
  \caption{Spectroscopic parameters of the different groups of variable stars described in the text.} 
  \begin{tabular}{ccc}
    \hline\hline\noalign{\smallskip}
 Class & $T_{\rm eff}$ & $\log g$   \\
       &   [K]         &          \\
\hline 
 $\delta$ Sct &$6000-9000$ & $ 3.25-4.4$  \\
 $\gamma$ Dor &  $6900-7700$ & $3.9-4.1$         \\
roAp          &   $6500-8500$  & $ 3.25-4.4$  \\
 ELMV         & $7000-10\,000$ & $6-7$           \\
 pre-ELMV     &  $6000-12000$& $ 3-5.5$       \\  

\hline\hline
\end{tabular}
\label{resumenspectro}
\end{table}

\begin{table}[h]
  \centering
  \caption{Period and frequency ranges usually observed of the different variable stars described in the text.} 
  \begin{tabular}{ccc}
    \hline\hline\noalign{\smallskip}
 Class & Period range & Frequency range  \\
       &      [s]        &  [c/d] \\
\hline 
 $\delta$ Sct & $700-28\,800$& $3-123.42$ \\
 $\gamma$ Dor & $26\,000-260\, 000$&$0.33-3.32$       \\
roAp           & $339-1\,272$& $67.92-254.86$\\
 ELMV          &  $1\,000-6\,300$& $13.71-86.4$         \\
 pre-ELMV      &   $320-6\,000$& $14.4-270$        \\  

\hline\hline
\end{tabular}
\label{resumenperiod}
\end{table}

In Fig. \ref{HRcompleto} we display the position of the different classes of variable stars analyzed in this work in the $T_{\rm eff}-\log g$ plane. $\delta$ Sct stars, depicted with white circles, are from \cite{2011A&A...534A.125U}, \cite{2015AJ....149...68B} and \citet{2016MNRAS.460.1970B}. SX Phe are those from \cite{2012MNRAS.426.2413B}, depicted by magenta circles. The error bars for the $\delta$ Sct stars and the SX Phe stars are depicted outside the $\delta$ Sct and SX Phe region in black and magenta, respectively. With blue diamonds we show the position of the known pre-ELMVs \citep{2013Natur.498..463M, 2014MNRAS.444..208M, 2016ApJ...822L..27G}, while the ELMVs are depicted with light green triangles \citep{2012ApJ...750L..28H, 2013ApJ...765..102H, 2013MNRAS.436.3573H,2015MNRAS.446L..26K,2015ASPC..493..217B}. The new pulsating J0757+1448 reported for the first time in this article is depicted with a cyan diamond and the cyan squares indicate the position of the two stars reported by \citet{2016A&A...587L...5C}, J1730+0706 and J1458+0707. We have also included theoretical evolutionary tracks for low mass He--core WDs from \cite{2013A&A...557A..19A} (dotted black lines in Fig. \ref{HRcompleto}) and \cite{2001MNRAS.325..607S} (dashed black line in Fig. \ref{HRcompleto}). Black numbers correspond to the values of the stellar mass of low mass He--core WD evolutionary tracks. In the same figure we illustrate MS evolutionary tracks (with red lines), for different values of metallicity ($Z$), mass ($M_{\rm \sun}$) and overshooting parameter ($\alpha_{\rm OV}$) from \cite{2017A&A...597A..29S}. The red numbers indicate different combinations of these parameters which are summarized in Table \ref{msmodels}. Up and down triangles show the position of two sets of models (see Section \ref{asteroseismic-comparison}). The obvious differences in the MS evolutionary tracks are due to the different values of stellar mass, metallicity and core overshooting parameter adopted. The occurrence of the overshooting phenomena extends the track of the stellar model toward lower values of $\log g$ and $T_{\rm eff}$, resulting ultimately in a notable broadening of the MS. This is because extra mixing leads to a larger amount of H available for burning in the stellar core which of course prolongs the MS lifetime. Note that unlike metallicity, overshooting does not change the location of the ZAMS in the HR diagram.

From Fig. \ref{HRcompleto} we note that the region where pre-ELMVs are located overlaps with the region occupied by the $\delta$ Sct stars. It is important to mention that the percentage of pre-ELM and ELM WDs "polluting" this region of the diagram ($6\,000 <T_{\rm eff}<12\,000$ K and $3<\log g <7$) is approximately $7 \%$ in a sample of subdwarf A-type (sdA) stars commonly observed in this region \citep{2018MNRAS.475.2480P}. Note that the physical nature of sdAs stars is still an open issue. Possible explanations include metal poor main sequence A-F stars \citep[see, e.g.,][]{2017ApJ...839...23B}, and binary byproducts such as ELMs, pre-ELMs and blue stragglers \citep[e.g.,][]{2018MNRAS.475.2480P}.

\begin{table}[h]
  \centering
  \caption{Stellar parameters of the MS evolutionary sequences.} 
  \begin{tabular}{cccc}
    \hline\hline\noalign{\smallskip}
 Mass & Overshooting parameter & Metallicity & Model   \\
  $M_{\rm \sun}$  &  $\alpha_{\rm OV}$     &  $Z$  & number    \\
\hline 
 2.2 & 0.03 & 0.02  & 1      \\
 2.2 & 0.03 & 0.01  & 2      \\    
 1.8 & 0.00 & 0.015 & 4       \\
 1.8 & 0.02 & 0.015 & 3      \\
 1.5 & 0.02 & 0.015 &  5 \\
 1.5 & 0.00 & 0.015 &  6     \\
 1.2 & 0.00 & 0.015 &  7   \\
 1.2 & 0.03 & 0.01  &  8  \\
\hline\hline
\end{tabular}
\label{msmodels}
\end{table}

Finally, we show in Fig. \ref{HRcompleto} the location of the theoretical blue edge and the empirical red edge of the $\delta$ Sct instability strip extracted from \citet{2000ASPC..203..443P}, and also the blue edge of the pre-ELMV instability strip computed by \cite{2016A&A...588A..74C}. We note that both instability strips overlap for $3 \lesssim \log g \lesssim 4.4$. In addition, the observed period ranges showed by all the variable stars mentioned also overlap between 400 and $12\, 000$ s (or between 7.2c/d and 216 c/d in frequencies) (see Table \ref{resumenperiod}). However, the nature of the pulsation modes would not be the same for $\delta$ Sct and pre-ELM WD stars. For $\delta$ Sct pulsators most of those periods would correspond to radial modes or non-radial $p$ modes, while for pre-ELM WDs only $g$-modes can have long periods in that range. Thus, it would be possible to distinguish $\delta$ Sct and pre-ELMVs by analyzing the pulsational properties of these two classes of variable stars. This is the main goal of the present study.

Among all types of variable stars described in this section, we have analyzed in particular the $\delta$ Sct stars and pre-ELMV classes, since they are found in the same region of the $\log g -T_{\rm eff}$ plane and they share a common range of observed periods. $\gamma$ Dor stars are found within the $\delta$ Sct instability strip, but they exhibit pulsation periods that are too long as compared to the periods usually observed in ELMVs and pre-ELMVs (see Table \ref{resumenperiod}). This is also the case for hybrid $\delta$ Sct-$\gamma$ Dor stars. The period ranges of roAp stars partially overlaps with that of pre-ELMV WDs, but roAp stars show an overabundance of rare earth elements and strong magnetic fields and both characteristics, easily detected, are not observed in pre-ELMV stars. Finally, ELMV stars are characterized by surface gravities ($\log g \gtrsim 6$) larger than for pre-ELMVs and $\delta$ Sct stars. Therefore, we exclude $\gamma$ Doradus, hybrid $\delta$ Sct-$\gamma$ Dor, roAp and ELMV stars from our analysis.

\section{Numerical tools}
\label{numerical-tools}

The pulsation analysis presented in this work has made use of full stellar evolution models computed with the {\tt LPCODE} stellar evolution code. {\tt LPCODE} computes the complete evolution of low-- and intermediate--mass stars, from the ZAMS, through the core H-burning phase, the He-burning phase, and thermally pulsing asymptotic giant branch (AGB) to the WD stage. Details of the
{\tt LPCODE} can be found in \cite{Althausetal2005,Althausetal2009,Althausetal2013}, \citet{2010ApJ...717..183R} and \citet{2015MNRAS.450.3708R} and references therein. Here, we comment on the main input physics relevant for this work. Further details can be found in those papers.

To account for convection we adopted the mixing length theory with the free parameter $\alpha = 1.66$ for the MS phase. Note that the calibration of $\alpha$ depends on the effective temperature and surface gravity and therefore change with the evolution as shown in \citet{2014MNRAS.445.4366T}. Although a non-constant treatment of $\alpha$ should be used during the evolution in the MS, we selected $\alpha = 1.66$ with which we have reproduced
the present luminosity and effective temperature of the Sun, $L_{\sun}= 3.842 \times 10^{33}$ erg s$^{-1}$ and $\log T_{\rm eff} = 3.7641$, when $Z= 0.0164$ and $X= 0.714$ are adopted, according to the $Z/X$ value of \citet{Grevessenoels1993}. For the WD phase we adopted the MLT in its ML2 flavor, with $\alpha = 1.0$. We include radiative opacities from the OPAL project \cite{1996ApJ...464..943I} complemented at low temperatures with the molecular opacities from \cite{2005ApJ...623..585F}. For the WD regime an updated version of the \cite{1979A&A....72..134M} equation of state was employed. In the present work, we considered pre-WD stellar models without element diffusion (i.e., radiative levitation, thermal and chemical diffusion and gravitational settling). In this way, it is assumed that pre-ELMV stars preserve the outer H/He homogeneous chemical structure resulting from the previous evolution. The presence of He in the envelopes of pre-ELMVs plays a fundamental role for exciting pulsation modes \citep{2016A&A...588A..74C,2016A&A...595L..12I}, and has been observed in the atmosphere of three pre-ELMVs reported in \citet{2016ApJ...822L..27G}.

We treated the extra-mixing (overshooting) as a time-dependent diffusion process by assuming that the mixing velocities decay exponentially beyond each convective boundary with a diffusion coefficient given by $D= D_0 \exp(-2 z / f H_{\rm p})$, where $D_0$ is the diffusive coefficient near the edge of the convective zone, $z$ is the geometric distance of the considered layer to this edge, $H_{\rm P}$ is the pressure scale height at the convective boundary, and $f$ is the overshooting parameter also named as $\alpha_{\rm OV}$ \citep{1997A&A...324L..81H,2000A&A...360..952H}. In this study we selected a $\delta$ Sct model with extreme overshooting parameter value of $\alpha_{\rm OV}= 0.03$. Finally, the metallicity is taken to be $Z = 0.01$ for this selected model. In the sake of simplicity, the impact of stellar rotation on the equilibrium models and on the pulsation spectra has been neglected in this work. Nevertheless, we include a brief discussion about the effect of stellar rotation in the next section.  

The pulsation computations employed in this work were carried out with the adiabatic and non-adiabatic versions of the {\tt LP-PUL} pulsation code described in detail in \cite{2006A&A...454..863C} and \cite{2006A&A...458..259C}, respectively. Briefly, the adiabatic version of the {\tt LP-PUL} pulsation code is based on the general Newton-Raphson technique to solve the fourth order system of equations and boundary conditions that describe linear, adiabatic, non-radial stellar pulsations, following the dimensionless formulation of \cite{1971AcA....21..289D}. For the non-adiabatic computations, the code solves the sixth-order complex system of linearized equations and boundary conditions as given by \cite{1989nos..book.....U}.

\section{Asteroseismic comparison}
\label{asteroseismic-comparison}  

In this section we analyze and compare the pulsational properties of two groups of stellar models. The $\delta$ Sct models were taken from \citet{2017A&A...597A..29S}, while the models of pre-ELM stars are from \citet{2016A&A...588A..74C}. The pre-ELMV models employed in this work were computed considering a solar metallicity progenitor, typical of Population I stars. We note that some pre-ELM WDs actually could belong to the halo population, and so they would come from low-metallicity progenitors. In that case, the H content of the pre-ELM WDs would be larger than that characterizing pre-ELM WDs coming from solar metallicity progenitors \citep[see, e.g.,][]{2016A&A...595A..35I}.
It is interesting to note that pre-ELMVs and evolved $\delta$ Sct stars show low radial order $p$ and $g$ modes which, strictly speaking, are mixed modes \citep[e.g.,][]{2010A&A...509A..90L,2016A&A...588A..74C}. However, for simplicity, we call them conventional $p$ and $g$ modes. 

In our analysis, we consider non-radial $p$ and $g$ modes with pulsation periods in the ranges $\sim 400-50\,000$ s and $\sim 100-10\,000$ s for the $\delta$ Sct model and the pre-ELMV model respectively (frequencies in the ranges $\sim 1.72-216$ c/d and $\sim 8.64-864$ c/d respectively), with harmonic degrees $\ell= 1$ and $\ell= 2$. We selected two sets of two models, one for $\delta$ Sct stars and the other one for a pre-ELM WD star, both models being characterized by similar atmospheric parameters. The atmospheric parameters along with other parameters for the models are listed in Tables \ref{modelos} and \ref{modelos2}. The first set of stellar models is in a cool region  of the $T_{\rm }-\log g$ plane with an effective temperature typically of $\delta$ Sct stars and where pre-ELMV are predicted to lie. This cool set of stellar models is characterized by $\log g\sim 4.08$ and $T_{\rm eff}\sim 7\,650$ K (Table \ref{modelos}), and is indicated with an orange triangle up symbol in Fig. \ref{HRcompleto}. The second set of stellar models is in a hot region of the $T_{\rm eff}-\log g$  diagram in which $\delta$ Sct stars and pre-ELMV WDs can be confused if we consider the external uncertainties in $T_{\rm eff}$ and $\log g$ of the pulsating objects usually finds in this region. This hot set is characterized by $\log g\sim 4.23$ and $T_{\rm eff}\sim 9\, 670$K  (Table \ref{modelos2}), and is indicated as a triangle down symbol in Fig. \ref{HRcompleto}. Note that there are usually no $\delta$ Sct stars at such high effective temperature and this is consistence with the non-adiabatic analysis for this model (not included) which do not predict any unstable period. Nevertheless, the next adiabatic analysis provides tools to distinguish between the two kinds of stars in this region of the $T_{\rm eff}-\log g$  diagram.

\subsection{Chemical profiles and propagation diagrams}

\begin{table}
  \centering
  \caption{Parameters characterizing our set of cool models. The position of the models in the $T_{\rm eff} - \log g$ plane in Fig. \ref{HRcompleto} is depicted as an orange up-triangle symbol.}

  \begin{tabular}{lcc}
    \hline\hline\noalign{\smallskip}
    & $\delta$ Sct &  pre-ELMV \\
\hline   
\hline
$T_{\rm eff}$ [K] & $7\, 670$ & $7\, 648$\\
$\log g$  &   4.071 & 4.088 \\
Stellar mass [$M_{\sun}$]& 1.55  & 0.176\\
Stellar radius [$R_{\sun}$] &  1.90 &  0.63 \\
Age [Myr] & $1\,376.01$ & $9\,244.51$ \\
Luminosity [$L_{\sun}$]&  11.18  & 1.21 \\  
Asymp. period spacing [s] & $3\,015.9281$ & 105.217 \\
Asymp. frequency spacing [mHz]& $0.624\times 10^{-1}$ & 0.122 \\
\hline\hline
  \end{tabular}
  \label{modelos}
\end{table}

\begin{table}
  \centering
  \caption{Parameters characterizing our set of hot models. The position of the models in the $T_{\rm eff} - \log g$ plane in Fig. \ref{HRcompleto} is depicted as an orange down-triangle symbol.}
  \begin{tabular}{lcc}
    \hline\hline\noalign{\smallskip}
    & $\delta$ Sct &  pre-ELMV \\
   \hline\hline

$T_{\rm eff}$ [K] & $9\, 675$ &  $9\, 669$\\
$\log g$  &   4.226 & 4.232 \\ 
Stellar mass [$M_{\odot}$]& 2.20  & 0.192\\
Stellar radius [$R_{\odot}$] &  1.89 &  0.56 \\
Age [Myr] & 190.81 & 9405.73 \\
Luminosity [$L_{\odot}$]&  28.20 & 2.42 \\  
Asymp. period spacing [s] & 4195.96 & 90.25 \\
Asymp. frequency spacing [mHz]& $0.730\times 10^{-1}$ & 0.149 \\
           
    \hline\hline

  \end{tabular}
  \label{modelos2}
\end{table}

\begin{figure}[h!]
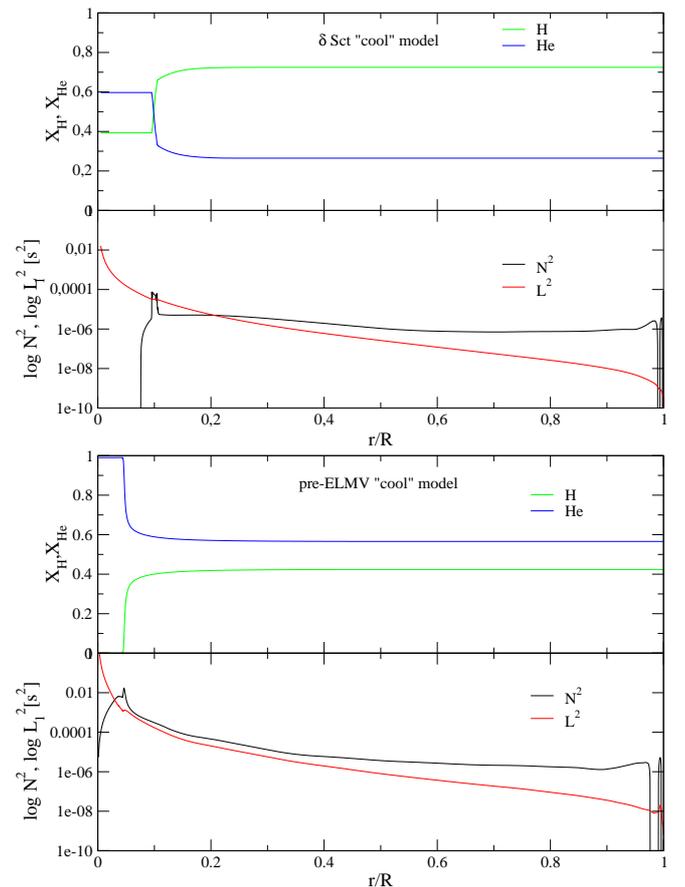

\begin{center}
\includegraphics[clip,width=8.6 cm]{deltasct/propa2.eps} 
\includegraphics[clip,width=8.6 cm]{pre-ELM/propa2.eps}
\caption{Abundances by mass of He and H as a function of the normalized radius (top panels), and propagation diagrams (the run of the logarithm of the squared Brunt-V\"ais\"al\"a and Lamb frequencies, lower panels) corresponding to the $\delta$ Sct model (upper panel), and to the pre-ELMV model (lower panel) indicated as a traingle up symbol in Fig. \ref{HRcompleto}.}
\label{propa} 
\end{center}
\end{figure}

\begin{figure}[h!]
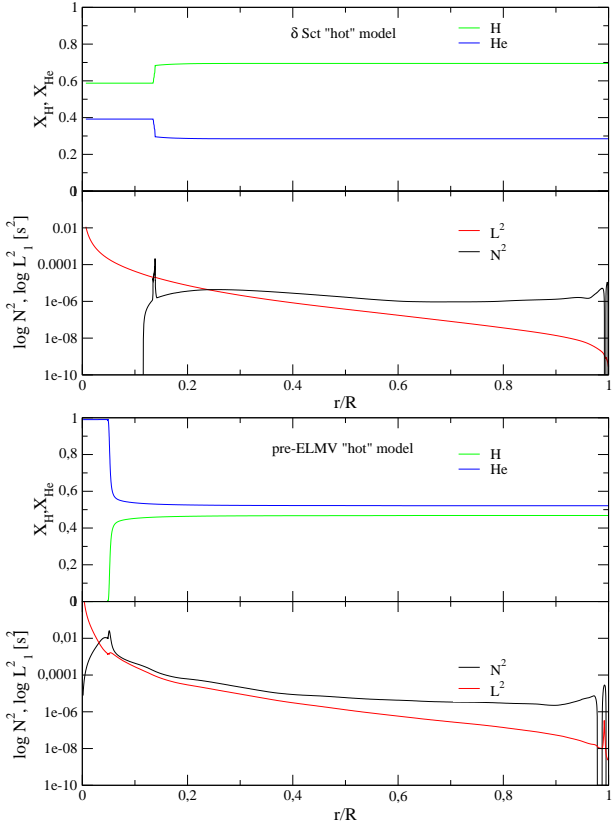

\begin{center}
\includegraphics[clip,width=8.0 cm]{deltasct/propaotro2.eps} 
\includegraphics[clip,width=8.0 cm]{pre-ELM/propaotro.eps}
\caption{Same as in Fig. \ref{propa}, but for the set of hot models, indicated as a triangle down symbol in Fig. \ref{HRcompleto}.}
\label{propa2} 
\end{center}
\end{figure}

We begin by analyzing the internal structure of the $\delta$ Sct and pre-ELMV models. In Figs. \ref{propa} and \ref{propa2} we show the inner chemical abundances of H and He and the propagation diagram as a function of the radius for our ``cool'' and ``hot'' models, respectively. Top panels correspond to the $\delta$ Sct models, while bottom panels correspond to the pre-ELMV models.  The $\delta$ Sct models show a H--dominated envelope and a H/He core, consistent with stars in the H--burning stage. The specific H/He ratio at the stellar core depends on the stellar mass and the age of the model. Since our cool $\delta$ Sct is older than the hot $\delta$ Sct model, the He abundances in the former is higher, leading to a He-dominated core star. The pre-ELM WD models, on the other hand, have a pure He--core since those models already depleted all the H in the core but did not reach the core temperature necessary to start the He burning in the core. These objects are consider as {\bf He--core} stars.

As shown in Figs. \ref{propa} and \ref{propa2}, the presence of internal density gradients associated to the He/H chemical transition regions leave strong signatures in the Brunt-V\"ais\"al\"a (B-V) frequency. For instance, considering the cool model showed in Fig. \ref{propa}, the bump in the B-V frequency for the $\delta$ Sct model (at r/R $\sim 0.1$) is higher and wider than for the pre-ELM model, indicating a wide chemical transition region for $\delta$ Sct stars. In addition, the B-V frequency is zero in convective regions. This can be seen for the $\delta$ Sct model,  at the convective core region and the outer convective layers. Note that, for the $\delta$ Sct model, the B-V frequency takes non-null values in an external layer of the core  (at $0.08 \lesssim r/R \lesssim 0.1$) with overshooting. In the case of the pre-ELMV model, the B-V frequency is zero only at the outer convective zone. In addition, the B-V and the Lamb frequencies take both higher values for the pre-ELMV model than for the $\delta$ Sct model, especially in the inner region. We note that the radius for the $\delta$ Sct model is approximately three times larger than for the pre-ELM model and the density for $\delta$ Sct star is much lower than for pre-ELM WDs. This increase the value of B-V frequency for pre-ELM star leading to a significant reduction of the mean period spacings of $g$ modes compared with mean period spacing of theoretical $g$ modes for the $\delta$ Sct model, which are usually associated with mixed modes.

\subsection{Periods and period spacings}  

In this section we analyze and compare the period ranges of dipole ($\ell= 1$) $p$ and $g$ modes for both, the pre-ELMV and $\delta$ Sct models. It is important to mention that not all modes are excited in $\delta$ Sct stars, which complicates mode identification and often hampers unique modeling. This mode selection mechanism is unfortunately still unclear \citep[see review e.g.,][]{2014IAUS..301..265S}. In addition, modes with $\ell=0,2,3..$ are also excited in these stars. We have explored the case of $\ell= 2$ and we have found similar results, thus we restrict ourselves to show the results for $\ell= 1$ only and we focus our analysis in the adiabatic pulsation spectra. 
 
\begin{figure}[h!] 
\begin{center}
\includegraphics[clip,width=8.0 cm]{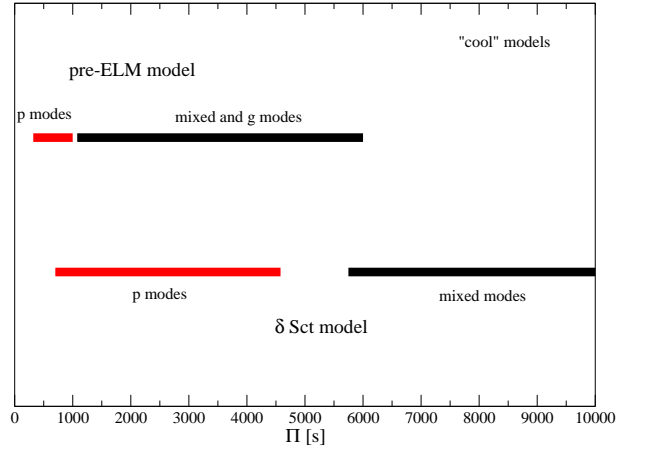} 
\caption{Adiabatic periods in the usually observed ranges computed for the cool $\delta$ Sct and pre-ELMV selected models, depicted with a triangle up symbol in Fig. \ref{HRcompleto}. The upper strips represent $p$ and mixed $g$ modes (in red and black, respectively) for the pre-ELMV model. The bottom strips represent $p$ and mixed modes for the $\delta$ Sct model.}
\label{superposteo} 
\end{center}
\end{figure}

\begin{figure}[h!] 
\begin{center}
\includegraphics[clip,width=8.0 cm]{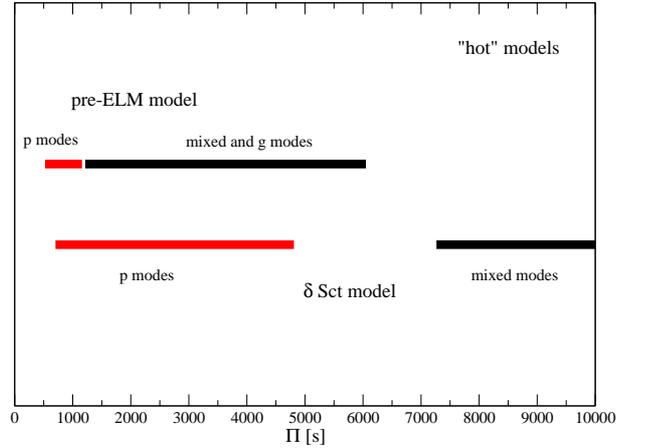} 
\caption{Same as Fig. \ref{superposteo}, but for the hot models depicted with a triangle down symbol in Fig. \ref{HRcompleto}.  }
\label{superposteootro} 
\end{center}
\end{figure}

The theoretical adiabatic periods in the usually observed period range for $p$ and $g$ dipole modes of the cool and hot models are depicted in Figs. \ref{superposteo} and \ref{superposteootro}. From these figures, we can see that the $\delta$ Sct and pre-ELMV models share a common range of periods. However, even if the period values are the same, the properties of the modes are not. For instance, in Fig. \ref{superposteo} we can see that the period range of $p$ modes for the $\delta$ Sct model overlaps with the period range of $p$ and $g$ modes for the pre-ELM WD model, while the periods of $g$ modes are much longer. Periods in the range $4\,579-5\,748$ s $15-19$ c/d) belong exclusively to the $g$ modes of the pre-ELMV model. Depending on the period value itself, the mode can be associated with a $p$ mode or a $g$ mode. For example, if an observed mode has a period of $\sim 2\,000$ s ($\sim 43.2$ c/d), then it can be a $g$ mode if it belongs to a pre-ELMV star, but it could alternatively be a $p$ mode if the object is actually a $\delta$ Sct star. However, for 
periods shorter than $\sim 1\,000$ s or larger than $\sim 5\,700$ s (frequencies larger that 86.4 c/d or shorter than 15.15 c/d), the modes will correspond to $p$ modes or $g$ modes, respectively, for both types of stars. 
Furthermore, by taking into account the usual range of observed periods in $\delta$ Sct ---which extends from $\sim 700$ s onward --- modes with periods below $700$ s belong exclusively to pre-ELMV stars (equivalently modes with frequencies above 123.42 c/d belong exclusively to pre-ELMV stars). The values of the periods and frequencies for dipole $g$ and $p$ modes for the cool models depicted in Fig. \ref{superposteo} are listed in Tables \ref{modosgms}, \ref{modosgpreelm}, \ref{modospms}, and \ref{modosppreelm} of the Appendix, along with their corresponding radial orders.

In order to decipher the nature of the observed periods, we need additional information, such as an estimation of the period spacing for $g$-modes($\Delta \Pi= \Pi_{k+1}-\Pi_{k}$) and the frequency spacing for $p$-modes ($\Delta \nu=\nu_{k+1}-\nu_{k}$, $k$ being the radial order). For high radial orders and for a given small value of $\ell$, the period spacing for $g$ modes tends to a constant value with no rotation or abrupt transitions in the stellar interior. Similarly, for $k>>l$ the large frequency separation (for $p$ modes) will also be constant, allowing to distinguish between $p$ and $g$ modes. Despite some stars showing high radial orders \citep[e.g.,][]{2011Natur.477..570A}, $\delta$ Sct and pre-ELMV stars rather exhibit pulsation in low to intermediate radial order. This means that is not completely adequate to use the asymptotic period spacing or the asymptotic frequency spacing to characterize the nature of the star. For this reason we focus on period spacings, frequency spacings, the mean period spacings ($\overline{\Delta \Pi}$) and mean frequency spacings ($\overline{\Delta \Pi}$), as defined below. The differences in these quantities along with the rate of the period change allow to distinguish between pre-ELMV and $\delta$ Sct stars.

We clarify that we use $\Delta \Pi$ and $\overline{\Delta \Pi}$ for $p$ modes only as math tools, to mimic the possible confusion between $p$ and $g$ modes, but they are not real seismic tools. The same applies for $\Delta \nu$ and $\overline{\Delta \nu}$ for $g$ modes. We assume no additional information other than the $T_{\rm eff}$ and $\log g$ values, and the observed pulsation modes, for which we have no prior knowledge concerning their nature. In order to properly distinguish between the well-known frequency and period spacings from  $\Delta \nu$ for $g$ modes and $\Delta \Pi$ for $p$ modes, we name the last two quantities as the frequency difference (of $g$ modes) and period difference (of $p$ modes) respectively. Note that these four quantities are computed for modes with consecutive radial order.

\begin{figure}
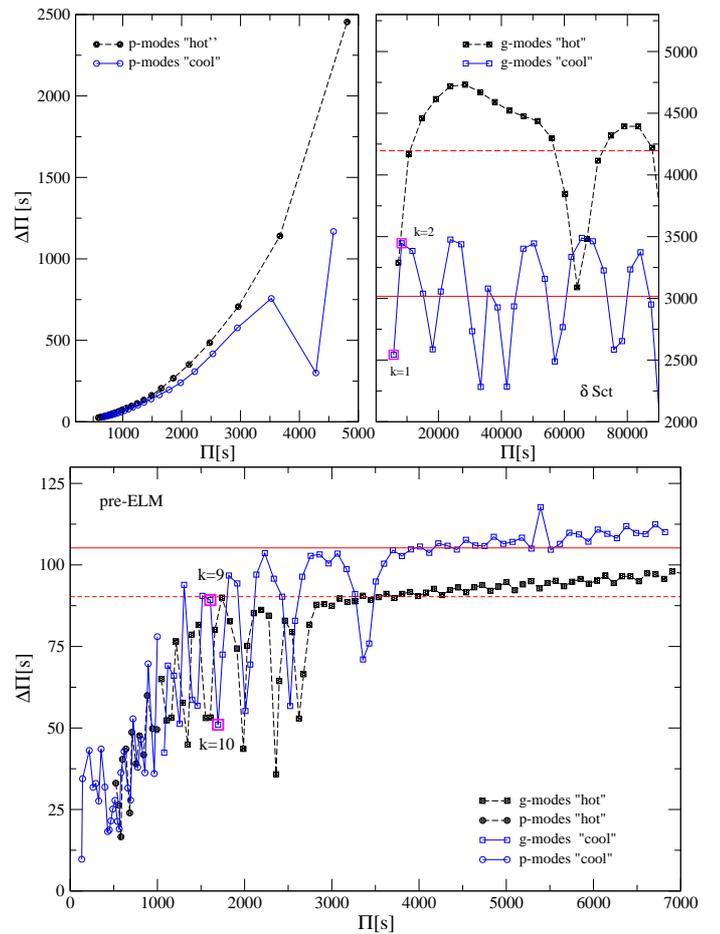

\begin{center}
\includegraphics[clip,width= 0.49\textwidth]{deltasct/DPvsPmodelo1y2todoslosmodos-ale.eps}\\ 
\includegraphics[clip,width=0.49\textwidth]{pre-ELM/DPvsPmodelo1y2todoslosmodos-ale.eps}
\caption{Period spacing vs. period for $p$ modes (circles) and $g$ modes (squares) corresponding to the $\delta$ Sct models (upper panel), and to the pre-ELMV models (lower panel). The asymptotic period spacing for $g$ modes is depicted as a reference with thin horizontal continuous and dashed lines for the cool and hot models, respectively. Selected periods with consecutive radial orders are marked with magenta squares. They correspond to $p$ modes for the $\delta$ Sct model, and to $g$ modes for the pre-ELMV model (see text for details). }
\label{DPvsP2} 
\end{center}
\end{figure}

In Fig. \ref{DPvsP2} we show the period spacing, $\Delta \Pi$, as a function of the periods, for the cool (blue) and hot (black) $\delta$ Sct (upper panel) 
and pre-ELM (bottom panel) template models. We include the asymptotic period spacing for $g$ modes, depicted with thin horizontal continuous and dashed red lines 
for the cool and hot models, respectively, as a reference. Let us consider the mean period spacing, defined as $\overline{\Delta \Pi}=(\Pi_j-\Pi_i)/N$, 
where $N$ is the number of consecutive radial orders modes with periods between $\Pi_i$ and $\Pi_j$. For theoretical $g$ modes, which are usually associated with 
mixed modes, the value of $\overline{\Delta \Pi}$ for the $\delta$ Sct model is larger than that for the pre-ELMV model in the usually range of observed periods 
for each kind of star. In particular, for the cool models, $\overline{\Delta \Pi}$ for the $\delta$ Sct model is $\sim 3\,000$ s (similar to the asymptotic 
period spacing) while for the pre-ELMV model the mean period spacing is more than an order of magnitude shorter, around $\sim 90$ s. Regarding the period spacing 
for $g$ modes for the $\delta$ Sct models, $\Delta \Pi$ ranges from $2\,300$ to $3\,500$ s approximately, while for the pre-ELMV model it ranges from  $\sim 50$ 
to 110 s. For $p-$modes, we also measured the $\overline{\Delta \Pi}$ value and it ranges from 50 to 300 s, depending on the range of periods ($\Pi_i$ and $\Pi_j$) 
considered for the $\delta$ Sct model, while it shows a more restricted range for the pre-ELM WD model, between 10 and 30 s. Besides, $\Delta \Pi$ ranges from 10 to 
$1\,200$ s for the $\delta$ Sct model, while it takes smaller values for pre-ELMV model, from 10 to 75 s. A similar trend is found for the hot models. We point 
out that the deviations from a constant value of $\Delta \Pi$  for $g$ modes in Fig. \ref{DPvsP2} are of course related with the chemical gradient and therefore with 
the profile of the B-V frequency. For a given star with a convective core, the numbers of minima in the period spacing increase as the star evolves on the MS due to 
the widening of the H/He chemical transition region \citep[e.g.,][]{2008MNRAS.386.1487M}.

\begin{figure}[h!]
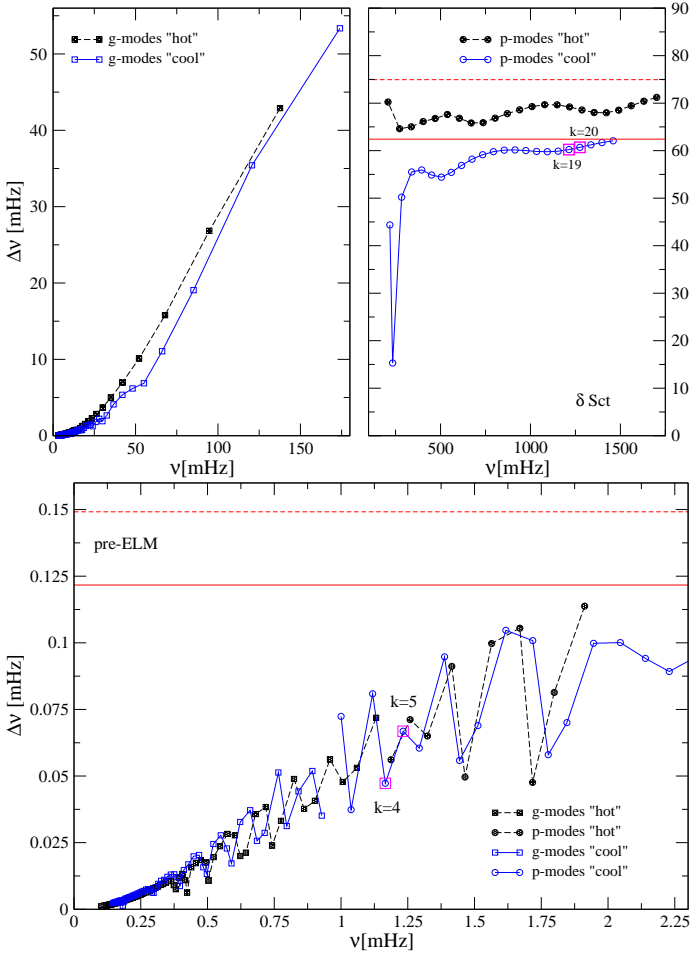

\begin{center}
\includegraphics[clip,width=0.49\textwidth]{deltasct/Dnvsnmodificar.eps}\\ 
\includegraphics[clip,width=0.49\textwidth]{pre-ELM/Dvvsvpreelm.eps}
\caption{Same as Fig. \ref{DPvsP2}, but for the frequency spacings. Horizontal continuous and dashed lines correspond to the asymptotic frequency spacing for $p$ modes for the cool and hot models, respectively.}
\label{Dnvsn} 
\end{center}
\end{figure}

In Fig. \ref{Dnvsn} we show the frequency spacing ($\Delta \nu$) versus frequency for the $\delta$ Sct models (upper panel) and the pre-ELMV models (lower panel). 
Horizontal continuous and dashed red lines correspond to the asymptotic frequency spacing for $p$ modes for the cool and hot model, respectively, included 
as a reference. The frequency spacing reflects the same trend that has been described for the period spacing. For example, considering the cool models, we can 
see that the mean frequency spacing (defined as $\overline{\Delta \nu}=(\nu_j-\nu_i)/N$) for $p$ modes of the $\delta$ Sct model is 6.56 c/d, while for the pre-ELMV 
model is 21.74 c/d. Furthermore, $\Delta \nu$ ranges between 1.32 and 5.46 c/d for the $\delta$ Sct model and from 2.59 to 8.64 c/d approximately for the pre-ELM model. 
For low order $g-$ modes, $\overline{\Delta \nu}$ is 2.06 c/d for the $\delta$ Sct model while for the pre-ELMV model it is 1.07 c/d. In addition, for low order $g$ 
modes $\Delta \nu$ ranges from 0.59 to 4.61 c/d for the $\delta$ Sct model and from 0.2 to 4.32 c/d for the pre-ELMV model.

Note that if the observed periods belong to modes with the same $\ell$ and are consecutive radial orders, then the $\Delta \Pi$ value and its behavior can also help to 
distinguish $\delta$ Sct from pre-ELMV stars, even if the observed modes have the same nature for both classes of pulsating stars. To illustrate this point, let us 
consider the $g$-mode periods corresponding to $k= 1$ and $k= 2$ for the $\delta$ Sct model, depicted with magenta squares in the upper panel of Fig. \ref{DPvsP2}, 
and the periods corresponding to $k= 9$ and $k= 10$ for the pre-ELMV model, depicted with magenta squares in the lower panel of the same figure. The period spacing 
for the selected periods of the $\delta$ Sct model is $2\,543$ s, while for the pre-ELMV model it is $89$ s, it is almost two orders of magnitude smaller.

By taking a closer look at the cool models in Fig. \ref{DPvsP2} we can note that for periods between $1\,000$ and $4\,600$ s ($18.78-86.4$ mHz) the low radial order $p$ modes corresponding to the $\delta$ Sct model overlap with the pre-ELMV $g$-modes. In particular, for the $\delta$ Sct model (see the magenta squares in upper panel of Fig. \ref{DPvsP2}), the $p$ modes from $k=1$ to $k=5$ have periods between $2\,500$ and $4\,600$ s (18.86 and 34.15 c/d), with a mean period difference of $\sim 510$ s (or mean frequency spacing of 3.05 c/d). In this same period interval, the mean period spacing for periods corresponding to high radial order $g$ modes for the pre-ELMV model (modes with radial order from $k=20$ to $k=41$) is $\sim 100$ s (or mean frequency difference of 0.72 c/d). This difference of $\sim 400$ s allows us to easily distinguish between consecutive low--radial order $p$ modes and high--radial order $g$ modes, and thus $\delta$ Sct stars from pre-ELMV stars. On the other hand, for $p$ modes with $k=6$ to $k=15$ for the $\delta$ Sct model, the periods are in the range [$1\,000-2\,500$] s (34.56 to 86.4 c/d) and the mean period difference is $\sim$ 132 s (mean frequency spacing of 5 c/d). In the same period range, $g$-mode periods of low radial order from $k=1$ to $k=19$ of the pre-ELMV model show a mean period spacing of 75 s (mean frequency difference of 2.29 c/d). Then, the difference in $\overline{\Delta \Pi}$ is $\sim 60$ s, which still could allow to distinguish between $p$ modes and $g$ modes. 

We point out that $\delta$ Sct stars are moderate to rapid rotators, showing projected rotational velocities ($V \sin i$) around 110 km/s \citep[see e.g.,][]{1997A&AS..122..131S}. On the other hand the only calculated $V \sin i$ values for pre-ELMVs are 30 km/s and 24 km/s for J0247-25B and WASP1628+10B respectively. Rotation induces to rotational splitting of the frequencies in the pulsation spectra. Considering rigid rotation and a first order perturbation theory we can estimate the maximum velocity ($V_{max}$) beyond which the difference between the frequency of the $m=0$ component and the $m=\pm 1$ component of the rotational multiplet ($\delta \nu$) is greater than the mean frequency spacing. Using this expression for the cool models: $\overline{\Delta \nu}=\delta \nu= (1-C_{kl})\frac{V_{max}}{2 \pi R}$ we calculate the velocity $V_{max}$ at which the rotational splitting mimic the frequency spacing. For the $\delta$ Sct model we obtain $\sim V_{max}=634$ km/s for $p-$ modes and $V_{max}=411$ km/s for low order $g$ modes. Note that the break up velocity for this $\delta$ Sct model is $\sim 394$ km/s, therefore the rotational splitting will not mimic the frequency spacing of consecutive radial order modes. On the the other hand, for the pre-ELMV model we obtained $V_{max}=693$ km/s for $p-$ modes and $V_{max}=50$ km/s for $g$ modes. The break up velocity for this model is $230$ km/s, therefore the splitting in $p$ modes will not be confused with the frequency spacing of consecutive radial order $p$ modes at any possible velocity. On the contrary, the rotational splitting of $g$ modes could be confused with the frequency separation if $ i \lesssim 37^{\circ}$ considering $V \sin i =30$ km/s for the pre-ELMV model. We conclude that this phenomena should be taken into account, in general cases, specially when a mode identification is required.

\subsection{Rate of period change}
\label{pdot}
  
In this section we analyze the values of the temporal rate of period change, $d\Pi/dt$. For the two cool template models, we estimated the $d\Pi/dt$ corresponding to a pulsation mode with a period of $\sim 966$ s (89.44 c/d).  From Tables \ref{modospms} and \ref{modosppreelm} we see that such a mode, emphasized in bold, corresponds to a $p$ mode with radial order $k= 16$ and $k= 2$ for the $\delta$ Sct and the pre-ELMV models, respectively. For the mode corresponding to the $\delta$ Sct model (with a period of 966.95 s and a frequency of 89.35 c/d) we obtain a rate of period change of $d\Pi/dt= 5.45 \times 10^{-5}$ s/yr. For the period corresponding to the pre-ELMV model (963.715 s or 89.65 c/d) the result is $d\Pi /dt= -1.42 \times 10^{-3}$ s/yr. Then, for the selected modes, the absolute value of the rate of period change for the pre-ELMV mode is two orders of magnitude higher than for the $\delta$ Sct model. These results are in agreement with \citet{1998A&A...332..958B} regarding $\delta$ Sct stars, and \citet{2017A&A...600A..73C} in connection with pre-ELMV stars.  Therefore, the rate of period change is a potential tool with which to distinguish the two kinds of stars.

We mention here that in the frame of the Kepler and K2 missions and the upcoming TESS mission, the rate of the period change is a powerful tool that could help to improve the classification of the stars observed by these mission. Given that is possible to measure $d\Pi /dt$ for $\delta$ Sct stars with approximately four years of observations, the 351 days of the continuous viewing zone of the TESS Satellite should be enough to measure  $d\Pi /dt$ for pre-ELMV WDs since these objects evolve more than one thousand times faster than $\delta$ Sct stars. It is important to note that unlike the measure of the $\Delta \Pi$ or $\Delta \nu$, the rate of the period change is independent of mode identification which makes it a powerful tool to distinguish between $\delta$ Sct and pre-ELM pulsators.

\section{Non-adiabatic analysis}
\label{no-adiabatico}

\begin{figure}[h!]
\begin{center}
\includegraphics[clip, width=0.48\textwidth]{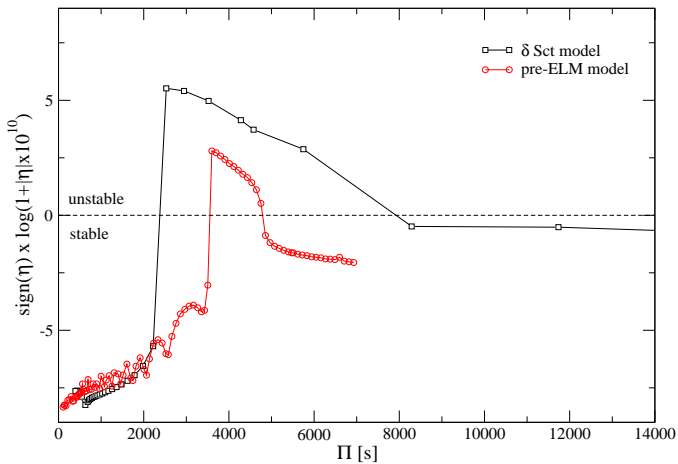}
\caption{Normalized growth rates $\eta$ for $p$ and $g$ modes
in terms of the pulsation periods corresponding to the selected models. Modes corresponding the $\delta$ Sct model are depicted with square symbols and continuous black line, while modes corresponding to the pre-ELM WD model are depicted with circle symbols and continuous red lines. A value of $\eta > 0$ ($ \eta < 0$) implies unstable (stable) modes. In frequencies, the band of excited modes for the $\delta$ Sct model is $10.16 \lesssim \nu \lesssim  34.56$ c/d) and for the pre-ELMV model is $17.28 \lesssim \nu \lesssim 24.68$ c/d. }
\label{inesta1} 
\end{center}
\end{figure}

In this section we present the results from non-adiabatic pulsation computations for the cool set of models. As mentioned, the calculations were carried out using the non-adiabatic version of the {\tt LP-PUL} pulsation code described in detail in \cite{2006A&A...458..259C}. Fig. \ref{inesta1} shows the normalized growth rate $\eta$ ($\equiv -\Im{(\sigma)}/\Re{(\sigma)}$) in terms of the pulsation periods $\Pi$ for $\ell = 1$ modes corresponding to the $\delta$ Sct (depicted with black squares) and pre-ELM WD template model (depicted with red circles). $\Im{(\sigma)}$ and $\Re{(\sigma)}$ are the imaginary and the real part, respectively, of the complex eigenfrequency ($\sigma$). If $\eta > 0$, then the mode is pulsationally unstable, and could reach observable amplitudes, while modes with $\eta < 0$ are stable. 

We can observe in Fig. \ref{inesta1} that the band of excited modes for the $\delta$ Sct model is $2\,500 \lesssim \Pi \lesssim  8\,500$ s ($10.16 \lesssim \nu \lesssim  34.56$ c/d) which correspond to low--order $g$ and $p$ modes. Overlapping this period interval, we found the unstable modes for the pre-ELMV model, spreading from $3\,500$ s to $5\,000$ s (17.28 c/d to 24.68 c/d), which are $g$ modes with relatively high radial orders. This means, for these particular models, that if we find excited modes in the ranges $2\,500 \lesssim \Pi \lesssim  3\,500$ s and  $5\,000 \lesssim \Pi \lesssim  8\,500$ s ( $24.68 \lesssim \nu \lesssim  34.56$ c/d and  $10.16 \lesssim \nu \lesssim  17.28$ s), they will correspond to a $\delta$ Sct star. In the case of having observed periods only in the interval $3\,500 \lesssim \Pi \lesssim 5\,000$ s ($17.28 \lesssim \nu \lesssim  24.68$ c/d), it is not possible to discern whether these observed periods correspond to a $\delta$ Sct star or a pre-ELMV star with this analysis only since the periods of excited modes of both stars overlap in this region and the period spacing analysis described previously should be done with the aim of properly classify the star. We consider the case of $\ell=1$ and these ranges might be increased if modes with $l=0,2,...$ are also taken into account.

\section{Testing the technique}
\label{testing}

In this section we present a new variable star, J075738.94$+$144827.5, with spectroscopic values of $T_{\rm eff}$ and $\log g$ in the region where the instability strips of pre-ELMV and $\delta$ Sct stars overlap {(see Table \ref{estrellas}). The atmospheric parameters were determined by fitting the optical spectra to DA local thermodynamic equilibrium (LTE) grids of synthetic nonmagnetic spectra derived from model atmospheres with metals in solar abundances from \citet{2010MmSAI..81..921K} \citep[see also][]{2018MNRAS.475.2480P}. We analyze the nature of this star in the frame of the previous discussion, considering the external uncertainties in $\log g$ and $T_{ \rm eff}$, which are approximately 0.25 dex and $5-10\%$ respectively.

\begin{table}[h]
  \centering
  \caption{Spectroscopic parameters and periods of J075738.94$+$144827.50.} 
  \begin{tabular}{ccccc}
    \hline\hline\noalign{\smallskip}
 $T_{\rm eff}$ [K] & $\log g$ & Period[s] & Freq. [c/d]&  Amp[mmag] \\

\hline 
 $8\, 180 \pm 250$ & $4.75 \pm 0.05$ & $2981.17$ &28.98 &1.87 \\
                 &                 & $2435.08$ & 35.48 &2.55  \\
                 &                 & $2055.73$ & 42.04&1.16  \\  
                  &                 & $802.90$ & 107.6& 0.49  \\
\hline\hline
\end{tabular}
\label{estrellas}
\end{table}

The variable star J075738.94$+$144827.5 was observed in the Southern Astrophysical Research (SOAR) telescope for 4.6 h during the night of 04 April 2016 as part of the project SO2016A-006 - Photometry of Extremely-Low Mass White Dwarf Candidates. It was observed with the Goodman HTS with an integration time of 10 s and the S8612 blue filter. The CCD region of interest was set to 800 $\times$ 800, resulting on a read out time of about 3.8 s. The light curve and Fourier transform are shown on Fig. \ref{0757}. Four pulsation periods were found above a $3 \sigma$ limit of detection, where $\sigma$ is the mean amplitude of the Fourier transform. Three periods are between $2\,000$ and $3\,000$ s (28.8 and 43.2 c/d), and one period is shorter (802.90 s or 107 c/d) and with smaller amplitude. The spectroscopic parameters along with the observed periods corresponding to this object are listed in Table \ref{estrellas}. The position of the star in the $T_{\rm eff}-\log g$ diagram on Fig. \ref{HRcompleto} is represented as a cyan full diamond.

\begin{figure*}
\begin{center}
\includegraphics[clip,width=18 cm]{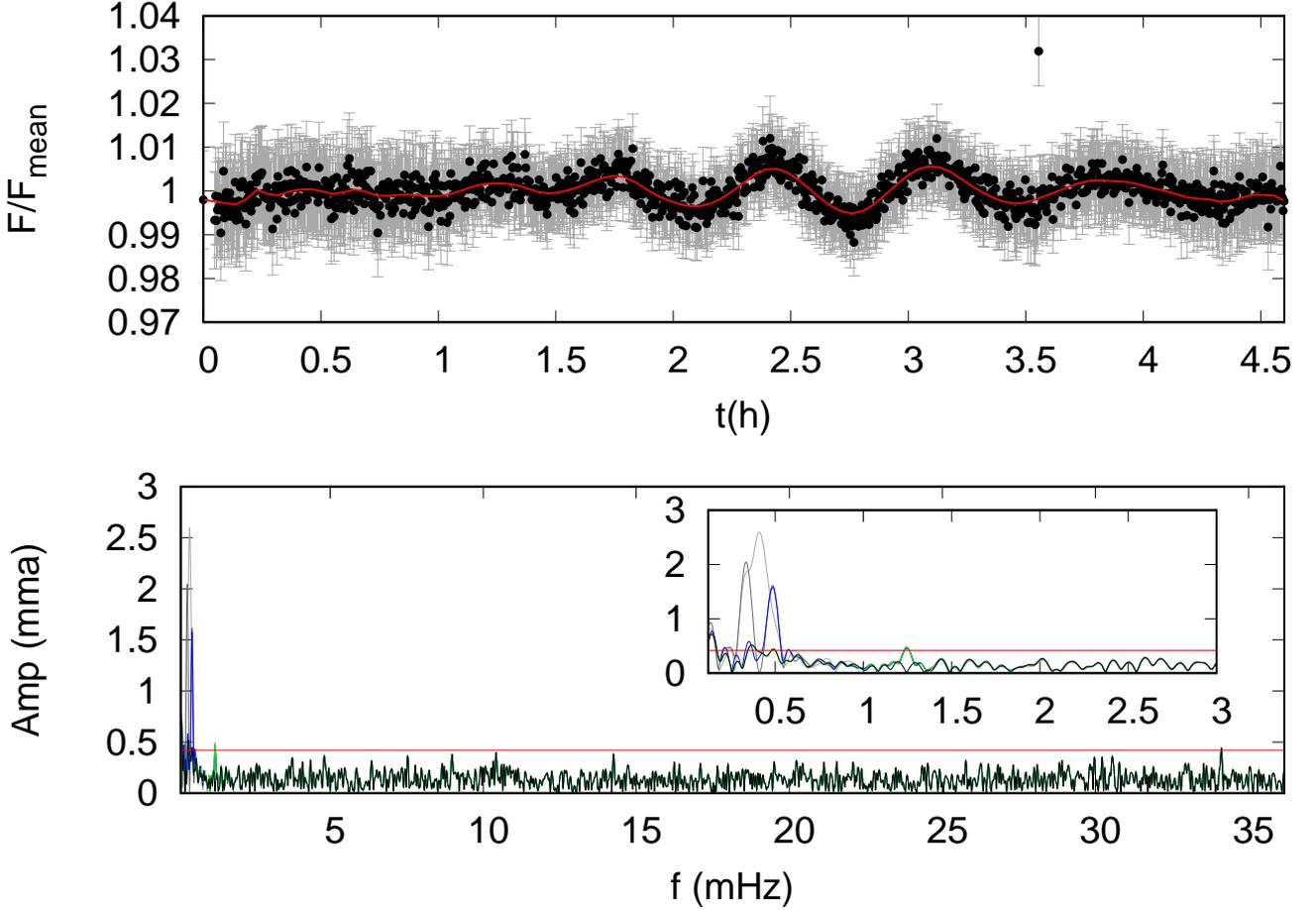} 
\caption{Top panel: Light curve for J075738.94$+$144827.5. Error bars are shown in gray and a smoothed light curve is shown in red. Bottom panel: Original Fourier transform in gray, and the subsequent Fourier transforms, subtracting periods above the 3˜$\sigma$ detection limit, in different colors. The final detection limit, above which no further periods were found, is shown in red.}
\label{0757} 
\end{center}
\end{figure*}

Regarding the nature of J075738.94$+$144827.5, and considering the $p$ and $g$ mode period ranges of the $\delta$ Sct and pre-ELMV models (Tables \ref{modosgpreelm} and \ref{modospms}), there are two possible scenarios. Specifically, the star could be a $\delta$ Sct star with four $p$ modes, or it could be a pre-ELMV star with three $g$ modes and one $p$ mode (the one with a period of $802.90$ s or 107.6 c/d) as well. If we assume that the periods between $2\,000$ and $3\,000$ s (28.8 and 43.2 c/d) have consecutive radial orders, then the observed period difference is between $380$ and $546$ s (frequency difference  $\sim 6.53$. These values are similar to the period difference values corresponding to $p$-modes of the $\delta$ Sct model in the range $2\,000-3\,000$ s, i.e., $\sim 300$ s (mean frequency spacing of $\sim 3.6$. Note that the mean period spacing for a pre-ELMV model in the same range of periods is of only $\sim 80$ s (mean frequency difference of $\sim 1.1$ c/d). Then we conclude that J075738.94+144827.5 is most likely a $\delta$ Sct star with four observed $p$ modes, if we consider the external uncertainties in the atmospheric parameters of this object. 

In addition, two interesting variable stars have recently been reported in \citet{2016A&A...587L...5C}: J145847.02$+$070754.46 and J173001.94+070600.25. Their location in the HR diagram ($\log g = 4.25$ and $T_{\rm eff} \approx 7950 K$) along with the lack of information about theirs masses make their classification difficult. Furthermore, the observed periods are compatible with those corresponding to $\delta$ Sct star and pre-ELMV WDs simultaneously. In the case of J173001.94$+$070600.25, the star shows only one period of $3\,367.1$ s (25.66 c/d), which is in the period range where both $\delta$ Sct and pre-ELMV stars show variability. However, as can be seen from Fig.  \ref{superposteo}, the observed period corresponds to a $p$ mode or a $g$ mixed mode, depending on the type of star. The detection of additional pulsation periods could allow us an estimate of the period spacing. A similar situation occurs for J145847.02$+$070754.46 which shows two periods of $1\,633.9$ s and $3\,278.7$ s (52.9 c/d and 26.35 c/d). According to those periods, these modes would correspond to $p$ modes if the star is a $\delta$ Sct star, and $g$ modes if the star is a pre-ELMV star. The difference between the two periods is $\sim 1\,645$ s ($\sim 26.55 c/d$), which is too large for the periods to have consecutive radial orders, either if the pulsating object is a $\delta$ Sct star or a pre-ELMV star. The rate of period change would help to classify these three stars. Unfortunately, these objects were observed during a short period and it is not possible to measure the rate of period change for them. We conclude that further observations are needed in order to allow a safe classification of these two stars on the basis of their pulsation properties.

\section{Conclusions}
\label{conclusions}

Pulsating pre-ELM stars and $\delta$ Sct are found roughly in the same region in the $\log g - T_{\rm eff}$ diagram, corresponding to an effective temperature in the range $6000 \lesssim T_{\rm eff} \lesssim 11000$ K and surface gravity in the range $3.5 \lesssim \log g \lesssim 5.5$, as can be seen from Fig. \ref{HRcompleto}. Specifically, theoretical models predict that pre-ELMV stars can be found in the $\delta$ Sct region ($6000 \lesssim T_{\rm eff} \lesssim 9000$ and $3.25 \lesssim \log g \lesssim 4.4$)} making it rather difficult to distinguish between the two groups from atmospheric parameters alone. In this paper, we performed a detailed analysis of the pulsational properties of $\delta$ Sct and pre-ELMV stars, aiming to use the differences in those properties to achieve a better classification. From the evolutionary and pulsation models of \citet{2017A&A...597A..29S} and \citet{2016A&A...588A..74C}, we have considered adiabatic and non-adiabatic pulsations for $\ell =1$ $p$, $g$ and mixed modes with periods between $\sim 400$ s and $\sim 10\ 000$ s ($\sim 8.64$ c/d and $\sim 216$ c/d) for the pre-ELMV model, and between $\sim 400$ s and $\sim 300\ 000$ ($\sim 0.28$ c/d and $\sim 216$ c/d) for the $\delta$ Sct model. Two sets of models for each class of stars were chosen to cover different regions in the $T_{\rm eff}-\log g$ diagram where pulsating objects are currently observed and can be confused with $\delta$ Sct stars or pre-ELM WDs. The cool set is characterized by atmosphere parameters of $T_{\rm eff}\sim 7\, 650$ K -- $\log g \sim 4.08$ representative of $\delta$ Sct stars and also predicted for pre-ELMV WDs. The hot set, with $T_{\rm eff}\sim 9\, 670$ K -- $\log g \sim 4.23$, is very close to the lower part of the classical instability strip, in a region where $\delta$ Sct and pre-ELMV can be confused if we consider the external uncertainties measured in these parameters. For each model, we analyzed the period range, period spacing and rates of period change. In addition, we analyze the pulsational stability of modes for the cool model. Finally, we compared the pulsational properties of both $\delta$ Sct and pre-ELMV stars, looking for differences and similarities in their pulsational properties. As mentioned above, we have assumed no additional information other than the $T_{\rm eff}$ and $\log g$ values, and the observed periods, for which we have no mode identification. Therefore we used $\Delta \Pi$ and $\overline{\Delta \Pi}$ for $p$ modes only as math tools and the same applies for $\Delta \nu$ and $\overline{\Delta \nu}$ for $g$ modes. Our main tools shown in this paper are the following:

\begin{enumerate}

\item The values of periods of $\delta$ Sct and pre-ELMV stars overlap, meaning that the observed mode may originate from either star. However, the nature of the mode (either $p$ or $g$ mode classification), can be different, depending on the nature of the star and the pulsation period. 

\item In the case of low radial orders and short periods ($\sim 1\, 000$ s and frequencies $\sim 86.4$ c/d), pulsation modes are $p$ modes for both kinds of stars. A similar situation is found for mixed modes with periods larger than $\sim 5\,700$ s (frequencies shorter than $15.15$ c/d), that correspond to $g$ mode pulsations. On the other hand, periods between 
$\sim 1\,000$ s and $\sim 5\,700$ s ($\sim 15.15$ c/d and $86.4$ c/d) are associated to $p$ modes if the star is a $\delta$ Sct star, while they are consistent with $g$ modes if the star is a pre-ELMV star. Once discovered, the different nature of the observed modes can be used to distinguish $\delta$ Sct from pre-ELMV stars if the periods are between $\sim 1\,000$ s and $\sim 5\,700$ s ($\sim 15.15$ c/d and $86.4$ c/d). 

\item The mean period difference for the $\delta$ Sct model is at least four times longer than that for the pre-ELMVs models in the period range between $2\,500$ and $4\,600$ s ($18.78$ and $34$ c/d). This difference, that amounts to $\sim 400$ s, allows us to distinguish between consecutive low--radial order $p$ modes and high--radial order $g$ modes, and thus $\delta$ Sct stars from pre-ELMV stars. A longer period difference for $\delta$ Sct stars is not surprising, since those stars have radii between $1.5$ and $3.5 R_{\sun}$ typical of MS stars, while pre-ELMV with the same $\log g$ have a radius of $\sim 0.6 R_{\sun}$.

\item The rate of period change shows a large difference between $\delta$ Sct and pre-ELMV stars. For the cool models, the rate of period change for a mode with a period of $\sim 965$ s is $d\Pi/dt= 5.453 \times 10^{-5}$ s/yr for the $\delta$ Sct model, while for the pre-ELM model it is $d\Pi /dt= -1.42 \times 10^{-3}$ s/yr. Note that $\delta$ Sct stars are evolving at MS rates of $\sim$ Gyrs, while pre-ELMV stars are in a giant star stage forced by the residual H burning before entering the WD cooling curve for the first time, evolving more than one thousand times faster than $\delta$ Sct stars.

\item Our preliminary non-adiabatic exploration indicates that for $\delta$ Sct and pre-ELMV stars, the ranges of unstable periods for the cool models overlap. 

\item We present a new pulsating star, J075738.94$+$144827.5, with four detected periods and spectroscopic parameters within the overlap region. We tried to determine the class of variable stars to which it belongs. The star could either be a $\delta$ Sct star with four $p$ modes, or it could be a pre-ELMV star with three $g$ modes and one $p$ mode as well. If we assume that the periods between $2\,000$ and $3\,000$ s (28.8 and 43 c/d) have consecutive radial orders, then the observed period difference is compatible with the period difference values corresponding to $p$-modes of the $\delta$ Sct stars, but not with the mean period spacing of the pre-ELMV model in the same range of periods. We conclude that J075738.94+144827.5 is most likely a $\delta$ Sct star, if we consider the external uncertainties in the atmospheric parameters of this object.

\end{enumerate}

The strong differences found in the period spacing of $g$ and $p$ modes, and in the rate of period change for $\delta$ Sct and pre-ELMV stars suggest that asteroseismology could be employed to discriminate between $\delta$ Sct and pre-ELMV stars. The Kepler/K2, TESS and PLATO missions will provide further observations needed to increase the number of pulsation modes, and to estimate the period and frequency spacing and/or the rate of period change in order to be able to distinguish between these two very different classes of pulsating stars. Finally, we will be able to confirm our results once the Gaia mission provides parallaxes for those objects with an unclear nature.

\begin{acknowledgements}
We thank our anonymous referee for their very relevant comments and suggestions that largely improved the content of the paper. JPSA especially thanks to Stellar Astrophysics Center and their funding provided by The Danish National Research Foundation (Grant agreement no.: DNRF106) for supporting this work. JPSA also thanks to the Henri Poincare Junior Fellowship during which this paper was finish. Part of this work was also supported by CNPq-Brazil and AGENCIA through the Programa de Modernizaci\'on Tecnol\'ogica BID 1728/OC-AR, by the PIP 112-200801-00940 grant from CONICET. Part of this work has been done with observations from the Southern Astrophysical Research (SOAR) telescope, which is a joint project of the Minist\'erio da Ci\^encia, Tecnologia, e Inova\c{c}\~ao (MCTI) da Rep\'ublica Federativa do Brasil, the U.S. National Optical Astronomy Observatory (NOAO), the University of North Carolina at Chapel Hill (UNC), and Michigan State University (MSU). This research has made use of NASA’s Astrophysics Data System.
\end{acknowledgements}

\bibliographystyle{aa} 
\bibliography{main} 

\appendix
\section{Tables}
Tables with the periods and frequencies of the cool set of models are presented below.

\begin{table}[h!]
  \centering
  \caption{Radial orders ($k$), periods in seconds and frequencies in units of mHz and \textbf{c/d} for the $g$ modes of the cool $\delta$ Sct model (triangle up symbol in Fig. \ref{HRcompleto}).}
\begin{tabular}{cccc}
\hline\hline
  k &  Period [s] & Freq. [mHz] & Freq. [c/d] \\ \hline
  1  &  5748.0217 & 0.17397290 & 15.0313 \\   
  2  &  8290.8955  & 0.12061423 &10.4211   \\
  3  &  11737.461 & 0.85197305E-01 & 7.36105  \\
  4  &  15121.630 & 0.66130438E-01&5.71367  \\ 
  5  &  18159.665 & 0.55067096E-01 & 4.7578   \\ 
  6  &  20746.138 & 0.48201743E-01 &4.16463 \\ 
  7  &  23801.457 & 0.42014234E-01& 3.63003 \\ 
  8  &  27277.711 & 0.36659967E-01 & 3.16742\\
  9  &  30715.686 & 0.32556655E-01& 2.81289\\
 10  &  33448.984 & 0.29896275E-01& 2.58304   \\
 11  &  35731.661 & 0.27986384E-01& 2.41802  \\
 12  &  38811.415 & 0.25765616E-01& 2.22615 \\
 13  &  41738.563  & 0.23958659E-01 & 2.07003   \\
 14  &  44023.330 & 0.22715229E-01& 1.9626 \\
 15  &  46958.650 & 0.21295331E-01& 1.83992  \\ 
 16  &  50361.060 & 0.19856612E-01& 1.71561  \\   
 \hline
\end{tabular}
\label{modosgms}
\end{table}

\begin{table}
  \centering
  \caption{Same as in Table \ref{modosgms}, but  for the $g$ modes of the cool pre-ELMV model (triangle down symbol in Fig. \ref{HRcompleto}).}
\begin{tabular}{cccc}
\hline\hline
  k & Period [s] & Freq. [mHz] & Freq. [c/d] \\ \hline
1  &  1077.7198  &0.92788493 &  80.1693 \\   
2 & 1120.1445 &  0.89274194 &  77.1329\\
  3 & 1189.2745  &  0.84084875 &72.6493\\
4 & 1255.3036   &    0.79662005 &68.828\\ 
5 & 1306.5457 &   0.76537695 &66.1286 \\ 
6 & 1400.4438  &    0.71405938 &61.6947\\ 
 7 & 1459.0320   & 0.68538595 &59.2173 \\ 
8 & 1515.8464 &    0.65969745 &56.9979\\
 9 & 1606.3661 &    0.62252309&53.786\\
10 & 1695.5839  &  0.58976735 &50.9559 \\
 11 &  1746.5640   &  0.57255274 &49.4686  \\
 12 & 1819.0768  &   0.54972939 &47.4966\\
  13 & 1915.8405  &  0.52196413 &45.0977 \\
 14 & 2010.1655 &  0.49747147&42.9815\\
 15 &  2065.3803 &    0.48417232&41.8325 \\ 
 16 & 2134.8242 &    0.46842265 &40.4717\\ 
 17 & 2231.8625  & 0.44805628 &38.7121\\ 
  18 &  2335.4907   & 0.42817555 &36.9944\\ 
 19 & 2431.2727   &  0.41130722 &35.5369\\
 20 & 2521.5297    &  0.39658465 &34.2649 \\
 21 &  2578.2539  &   0.38785940&33.5111\\
22 &  2661.1221 &   0.37578133 &32.4675\\  
  23 &  2757.5222  &    0.36264440&31.3325\\   
 24 & 2860.3180  &  0.34961148 &30.2064\\
  25 & 2963.5512  &   0.33743301 &29.1542 \\
  26 & 3064.0158   &  0.32636907 &28.1983 \\
 27 &3167.4945 &   0.31570694 &27.2771\\
 28 & 3266.2088 &    0.30616536&26.4527 \\ 
 29 &  3357.3064  &  0.29785783 &25.7349 \\
  30&   3428.3023 &    0.29168956 &25.202  \\
  31 & 3504.1919 &  0.28537250 &24.6562\\
  32 & 3599.1062   &  0.27784676&24.006\\
  33 &  3699.5019   & 0.27030666 &23.3545   \\
  34 &  3804.0303 &  0.26287909 &22.7128 \\
 35&  3906.7701  &  0.25596592&22.1155\\
 36 & 4011.5483 &   0.24928031&21.5378  \\
  37 &  4117.1873  &   0.24288426&20.9852  \\
 38 &  4220.9141 &  0.23691551 &20.4695   \\
39 & 4327.5460  &   0.23107784 &19.9651 \\
 40 & 4433.4207  &  0.22555946 &19.4883\\
 41 & 4538.1359   & 0.22035480  &19.0387 \\
 42 & 4645.8408   &  0.21524629 &18.5973 \\ 
 43 &  4751.8075 &   0.21044623  &18.1826\\
 44 & 4857.6097    &    0.20586256 &17.7865\\ 
45 & 4966.2464   &   0.20135932 &17.3974\\
46 &  5072.7264  &     0.1971326&17.0323\\
47 & 5179.7979   &     0.19305772 &16.6802\\
 48 &  5288.1439 &     0.18910227 &16.3384\\
49 & 5393.1616   &    0.18542000&16.0203\\   
 50 &  5471.1905  &    0.18277558 &15.7918 \\
 51 &  5510.9013  &    0.18145852 &15.678\\
   52 &5615.5779   &   0.17807606 &15.3858\\
 53 &  5722.0394   &    0.17476286&15.0995\\
  54 &  5831.9158   &     0.17147024&14.815 \\
 55 & 5941.2743  &   0.16831406  &14.5423\\
 56 &  6048.4160 &    0.16533254&14.2847\\ 
57 &6159.2673 &    0.16235697&14.0276\\
 58 & 6268.8071   &  0.15951998 &13.7825\\ 
59 &6377.0121   &   0.15681325 &13.5487\\
  60 & 6488.9050 &    0.15410921 &13.315  \\
 61 & 6598.7174 &    0.15154460  &13.0935 \\
 62 & 6708.2016   &   0.14907125  &12.8798\\
63 & 6820.6575   &    0.14661343&12.6674 \\ 

\hline
\end{tabular}
\label{modosgpreelm}
\end{table}

\begin{table}
  \centering
  \caption{Same as in Table \ref{modosgms}, but for $p$ modes of the cool $\delta$ Sct model (triangle up symbol in Fig. \ref{HRcompleto}). The rate of period change discussed in the text is calculated for the mode indicated in bold font style.}
\begin{tabular}{cccc}
\hline\hline
  k & Period [s] & Freq. [mHz] & Freq.[c/d]  \\ \hline
  27  & 440.66300  &   2.2693078   & 196.068  \\
26 & 480.06639 &   2.0830452 &179.975 \\
  25 &   397.14276 &       2.5179863 & 217.554  \\  
  24 &   631.11100 &       1.5845073 &136.901   \\
  23 &   685.04889 &        1.4597498 &126.122 \\
  22 &   715.47043 &       1.3976818  &120.76  \\
  21 & 748.51276   &       1.3359826  &115.429 \\
  20 & 784.46356   &       1.2747565  &110.139\\
  19 & 823.68257   &       1.2140599  &104.895 \\
  18 &  866.67120  &      1.1538401   &99.6918 \\
  17 & 914.11943   &     1.0939490    & 94.5172  \\
  16 &  \textbf{966.95122}  &      1.0341783   & 89.353 \\
  15 & 1026.3412   &    0.97433485    & 84.1825 \\
  14 &  1093.7161  &      0.91431409  & 78.9967 \\
  13 &  1170.7270  &      0.85417009  & 73.8003 \\
  12 &  1259.3439  &     0.79406425   & 68.6072  \\
  11 &  1361.8839  &     0.73427698   & 63.4415 \\ 
  10 &  1481.2000  &    0.67512828    & 58.3311 \\ 
   9 & 1620.9294   &     0.61693001   & 53.3028 \\ 
   8 &  1785.5962  &   0.56003704     & 48.3872 \\
   7 &  1981.7053  &   0.50461589     &43.5988 \\
   6 &  2221.3037  &     0.45018608   & 38.8961\\
   5 & 2529.6773   &     0.39530735   & 34.1546  \\
   4 &  2946.6472  &     0.33936875   & 29.3215 \\
   3 &  3523.0171  &    0.28384761    & 24.5244 \\
   2 & 4280.1302   &     0.23363775   & 20.1863  \\
   1 & 4579.9606   &     0.21834249   & 18.8648 \\ \hline
\end{tabular}
\label{modospms}
\end{table}

\begin{table}
  \centering
  \caption{Same as in Table \ref{modosgms}, but for $p$ modes of the cool pre-ELMV template model (triangle down symbol in Fig. \ref{HRcompleto}). The rate of period change discussed in the text is calculated for the mode indicated in bold font style.}
\begin{tabular}{cccc}
\hline\hline
  k & Period [s] & Freq. [mHz] & Freq. [c/d]   \\ \hline
  26 & 132.55850  &   7.5438392 &651.788\\
 25 & 142.28993 &      7.0279044&607.211\\
24 &219.42525 &     4.5573607 &393.756\\  
  23 &  262.53175 &     3.8090631 &329.103 \\
22 &294.33302   &      3.3975121&293.545\\
 21 &327.31241   &  3.0551851&263.968\\
 20 &354.92072 &      2.8175306&243.435\\
19 & 398.44561 &      2.5097528&216.843\\
18 & 430.31485  &      2.3238798&200.783\\
17 &448.53324   &   2.2294892 &192.628\\
 16 &467.23201  &      2.1402643 &184.919 \\
15 &488.73314  &     2.0461064&176.784\\
 14 &513.86912  &     1.9460208&168.136\\
 13 & 541.64878 &      1.8462148& 159.513\\
 12 &563.01488  &    1.7761520&153.46\\
11 &582.02471  &   1.7181401&148.447 \\
 10 &618.30639  &   1.6173212&139.737\\ 
 9 &661.11271   &     1.5126014 &130.689\\ 
 8 &692.68113   &   1.4436657&124.733\\ 
  7 &720.54634   &   1.3878358 &119.909\\
6 & 773.34419  &     1.2930853&111.723\\
5 &811.28450   &  1.2326132 &106.498\\
  4 & 857.77214  &   1.1658108 &100.726 \\
 3 & 894.03032  &     1.1185303&96.641\\
   2 & \textbf{963.71573}  &     1.0376504 &89.653\\
 1 & 999.72392  &     1.000276 &86.4238\\ \hline
\end{tabular}
\label{modosppreelm}
\end{table}

\end{document}